\documentclass[aps,prb,twocolumn,showpacs,amssymb,floatfix]{revtex4-1}
\usepackage{amsmath,graphicx}
\usepackage{mathrsfs}
\usepackage{bm}
\usepackage{color}

\newcommand{\la}{\langle}

\newcommand{\nn}{\nonumber}

\newcommand{\ra}{\rangle}

\newcommand{\alp}{\alpha}

\newcommand{\lam}{\lambda}

\newcommand{\sg}{\sigma}

\newcommand{\etal}{{\em et al.~}}

\begin{document}
\title{Tunable spin-charge conversion through topological phase transitions in zigzag nanoribbons}
\author{Hang Li$^{1}$}{}
\author{Aur\'elien Manchon$^{1}$}\email{Aurelien.Manchon@kaust.edu.sa}
\affiliation{$^{1}$King Abdullah University of Science and Technology (KAUST), Physical Science and Engineering Division (PSE).
Thuwal, 23955-6900, Saudi Arabia.}
\date{\today}


\begin{abstract}
We study spin-orbit torques and charge pumping in magnetic quasi-one dimensional zigzag nanoribbons with hexagonal lattice, in the presence of large intrinsic spin-orbit coupling. Such a system experiences topological phase transition from a trivial band insulator to a quantum spin Hall insulator either by tuning the magnetization direction or the intrinsic spin-orbit coupling. We find that spin-charge conversion efficiency (i.e. spin-orbit torque and charge pumping) is dramatically enhanced at the topological transition, displaying a substantial angular anisotropy.
\end{abstract}

\pacs{72.25.Dc,72.20.My,75.50.Pp} 
\maketitle

{\em Introduction} - Topological insulators, a new phase of matter, have attracted intense research interest due to their nontrivial physical properties and potential applications in spintronics\cite{Ortmann}. Similarly to conventional or band insulators they possess a band gap in the bulk. Yet, differently from conventional insulators they support time-reversal-symmetry-protected spin-polarized surface or edge states in the bulk band gap. These materials may experience a topological phase transition from topological to band insulators by structural design and manipulations such as doping with impurities\cite{Chen,Xu}, applying a strain or pressure\cite{Hirahara,Xi}, inducing lattice distortion or enhancing spin-orbit coupling via nonmagnetic substrates\cite{Kim}. Interestingly, even without structural manipulation, topological phase transition can also be driven by coupling topological insulators with magnetic substrates. For example, a transition from a band insulator to a quantum anomalous Hall insulator can be achieved by inducing magnetic exchange in silicene via a proximate magnetic layer\cite{Ezawa,Pan2014}. Among these studies, the influence of topological phase transitions on Hall conductivities and spin textures in momentum space has been confirmed \cite{Bernevig,Xu}. From a topological standpoint, a charge or spin current in a topological insulator is also a topological current. Hence, unlike semiconductors and metals, charge conductivities and spin-polarized edge states in topological insulators can be controlled not only by an electric field but also by topological phase transitions.\par

Besides topological phase transitions, charges flowing at the surface or edge of topological insulators are accompanied by a non-equilibrium spin polarization due to the large spin-momentum locking of surface states \cite{Theory1}. Such a magneto-electric effect can be used to excite and switch the magnetization of a ferromagnet deposited on the surface, as studied theoretically \cite{Theory2,Theory3} and demonstrated experimentally \cite{Fan2014,Wang2015,Mellnik}. This spin-orbit torque displays a larger electrical efficiency compared with spin torque in bilayers involving heavy metals\cite{miron,liu}. Alternatively, the spin-to-charge conversion present at the surface of topological insulators can be probed through charge pumping, i.e. the Onsager reciprocal of spin-orbit torques \cite{Brataas2011,Freimuth2015}. In fact, while a charge current creates a torque on the magnetization, a precessing magnetization induces a charge current along the interface. This effect was originally observed in magnetic bilayers involving heavy metals and attributed to the inverse spin Hall effect present in the bulk of the heavy metal\cite{Saitoh}. This observation has been recently extended to two dimensional systems such as hexagonal lattices \cite{pumpinggraphene}, semimetal surfaces \cite{Rojas-Sanchez2013} and more recently to the surface of topological insulators \cite{Shiomi,Jam,Sanchez}. In these systems, the spin-charge conversion is attributed to the spin-momentum locking induced by interfacial (Rashba or Dirac) spin-orbit coupling. While magneto-electric effects have been studied in topological insulators in the metallic regime \cite{Theory2,Theory3}, the influence of topological phase transitions on these mechanisms has been essentially overlooked. In particular, besides the emergence of quantized magneto-electric effect \cite{Theory1}, it is not clear how the topologically non-trivial edge states contribute to spin-orbit torque and charge pumping.

In this paper, we theoretically investigate both charge pumping and spin-orbit torque in quasi-one-dimensional zigzag nanoribbons with a hexagonal lattice in the presence of intrinsic spin-orbit coupling and magnetic exchange. Depending on the spin-orbit coupling strength, this system displays topological phase transitions between trivial (metallic) and non-trivial (quantum spin/anomalous Hall) phases \cite{Kane,Ezawa}. We demonstrate that spin-charge conversion efficiency is dramatically enhanced at the topological transition, resulting in large damping-like spin-orbit torque and DC charge pumping.

{\em Spin-orbit torque and charge pumping} - Let us first formulate the reciprocity relationship between spin-orbit torques and charge pumping (see also Ref. \onlinecite{Brataas2011,Freimuth2015}). We start from the definition of magnetization dynamics and charge current
 \begin{align}
 \partial_{t} {\bf m}=&\gamma {\bf m} \times \partial_{{\bf m}}F +{\hat \chi}\cdot {\bf E},\nn\\
 {\bf J}_{c}=&\hat{\bf{\sigma}} \cdot {\bf E}+\hat{\bf{\xi}} \cdot \partial_{{\bf m}}F, 
\end{align}
where $-\partial_{{\bf m}}F=-\partial_{{\bf m}}\Omega/M_{s}$ is the effective field that drives the dynamics of the magnetization in the absence of charge flow. $\Omega$ is the magnetic energy density and $M_s$ is the saturation magnetization. ${\bf E}$ is the electric field that drives the charge current through the conductivity tensor ${\rm\hat\sigma}$ in the absence of magnetization dynamics. ${\hat \chi}$ and ${\hat \xi}$ are the tensors accounting for current-driven torques and charge pumping, respectively. We can rewrite these two equations in a more compact form
\begin{equation}       
\left(                 
  \begin{array}{cc}   
   \partial_{t}n_{i}  \\  
   \partial_{t}m_{i}\\  
  \end{array}
\right)   
=\left(                 
  \begin{array}{cc}   
    L_{n_{i},f^{j}_e} & L_{n_{i},f^{j}_m} \\  
    L_{m_{i},f^{j}_e} & L_{m_{i},f^{j}_m}\\  
  \end{array}
\right)                
\left(                 
  \begin{array}{cc}   
   f^{j}_e\\  
   f^{j}_m\\  
  \end{array}
\right)   
\end{equation}
where we define the particle current $\partial_{{t}}n_{i}= S \rm J_{c,i}/e$ , the electric and magnetic forces $f^{j}_{e}=de E_{j}$, $f^{j}_{m}=\mu_{B}\partial_{m_{j}} F$. Onsager coefficients are then
\begin{equation}       
\left(                 
  \begin{array}{cc}  
    L_{n_{i},f^{j}_e} & L_{n_{i},f^{j}_m} \\ 
    L_{m_{i},f^{j}_e} & L_{m_{i},f^{j}_m}\\ 
  \end{array}
\right)   
=\left(                 
  \begin{array}{cc}   
   W\sigma_{ij}/e^2 & {\xi}^{ij}/\mu_{B} \\  
    { \chi}^{ij}/d&-(\gamma/\mu_{B})({\bf e}_{i}\times{\bf e}_{j})\cdot\bf m\\ 
  \end{array}
\right).                 
\end{equation}
Here, we consider a magnetic volume of width $ W$, thickness $d$ and section normal to the current flow $ S = W d$. Applying Onsager reciprocity principle \cite{Brataas2011,Onsager}
\begin{align}
 L_{n_{i},f^{j}_m}({\bf m})=-L_{m_{j},f^{i}_e}(-{\bf m}),
\end{align}
and we obtain ${ \xi}^{ij}({\rm\bf m})/\mu_{B}=-{ \chi}^{ji}(-{\rm\bf m})/d$. In two-dimensional magnets with interfacial inversion asymmetry, the spin-orbit torque ${\bf T}={\hat \chi}\cdot {\bf E}$ can be parsed into two components (see e.g. Refs. \onlinecite{Theory3,Hang-2015})
  \begin{align}\label{eq:torque}
{\bf T}=\tau_{\rm DL}{\rm\bf m}\times (({\rm\bf z}\times{\rm\bf E})\times{\rm\bf m})+\tau_{\rm FL}{\rm\bf m}\times({\rm\bf z}\times{\rm\bf E}),
\end{align}
referred to as the damping-like ($\tau_{\rm DL}$) and field-like torque ($\tau_{\rm FL}$). Hence, by definition
 \begin{align}
{ \chi}^{ij}({\bf m})=&\tau_{\rm DL}[{\bf m}\times (({\bf z}\times{\bf e}_{j})\times{\bf m})]\cdot {\bf e}_{i}\nn\\
&+\tau_{\rm FL}[{\bf m}\times({\bf z}\times{\bf e}_{j})]\cdot {\bf e}_{i}.
\end{align}
Then, applying Onsager reciprocity, we obtain the charge pumping coefficient 
\begin{align}
{ \xi}^{ij}({\bf m})=&-(\mu_{B}/d)\tau_{\rm DL}(-{\bf m})[{\bf m}\times (({\bf z}\times{\bf e}_{i})\times{\bf m})]\cdot {\bf e}_{j}\nn\\
&+(\mu_{B}/d)\tau_{\rm FL}(-{\bf m})[{\bf m}\times({\bf z}\times{\bf e}_{i})]\cdot {\bf e}_{j}.
\end{align}
And finally, the charge current induced by the magnetization dynamics reads
\begin{align}\label{eq:Jcp}
{\bf J}_{c}=&-\frac{\mu_{B}}{d\gamma}\tau_{\rm DL}(-{\bf m}){\bf z}\times({\bf m}\times{\partial_{t}\bf m})\nn\\
+&\frac{\mu_{B}}{d\gamma}\tau_{\rm FL}(-{\bf m}){\bf z}\times\partial_{t}\bf m.
\end{align}
This equation establishes the correspondance between the current-driven spin-orbit torque and the charge current pumped by a time-varying magnetization. In the following, we will compute the current-driven spin density $\delta {\bf S}$ from Kubo formula [Eq. (\ref{eq:TSP})]. The torque is simply ${\bf T}=(2J_{\rm ex}/\hbar) {\bf m}\times\delta{\bf S}$, so that that the conclusions drawn for spin-orbit torques equally apply to charge pumping.


\begin{figure}[tbh]
\begin{center}
\includegraphics[trim=0mm 0mm 0mm 0mm,clip,scale=0.39]{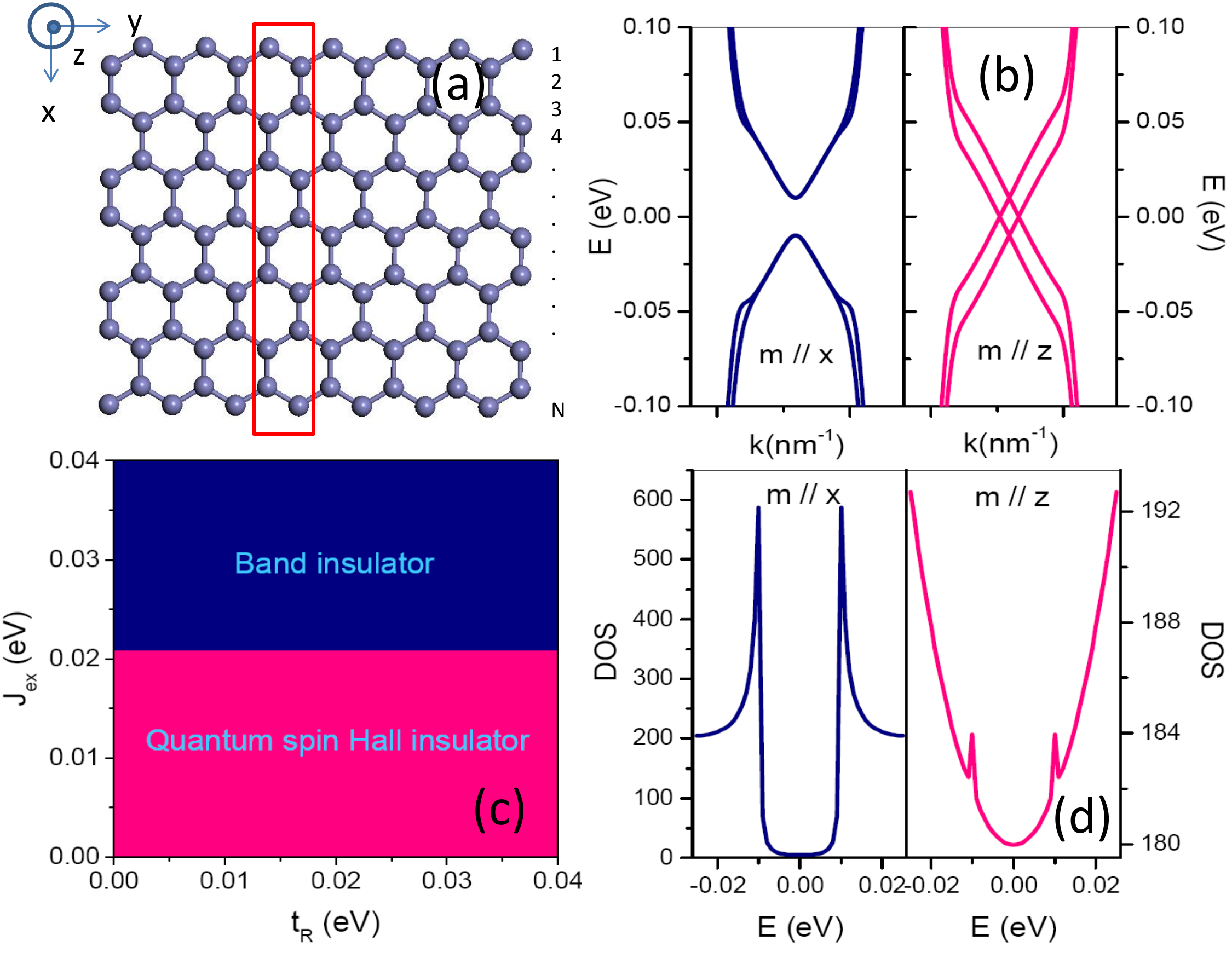}
\end{center}
\caption{(Color online) (a) Top view of zigzag silicene-like nanoribbons switched by a nonmagnetic topological insulator and a magnetic topological insulator. The super unit cell as indicated by the red rectangle. (b) Band structure for different magnetization direction ${\bf m}$. (c) Phase diagram for various Rashba and magnetization. (d) Density of states for different magnetization direction. The current is directed along the x axis. The parameters are ${t}_{\rm so}$=~36 meV and $J_{\rm ex}$= 10 meV.}
\label{fig1}
\end{figure}

{\em Model and method} - Let us now consider a single-layered zigzag nanoribbon with a hexagonal lattice (e.g. silicene, germanene, stanene etc.) deposited on top of a ferromagnetic layer. The ferromagnetic layer may be chosen as EuO\cite{Swartz} or YIG\cite{pumpinggraphene,Wang}, and induces a weak exchange coupling on the spin-polarized carriers as well as Rashba spin-orbit coupling.\par
 
 In a tight-binding representation, Hamiltonian for silicene-like material can be described by\cite{Qiao}
   
\begin{align}
H_{0}=\sum_{\langle i,j \rangle{\alp}}t\hat{c}_{i,{\alp}}^{+}\hat{c}_{j,{\alp}}+i\frac{{t}_{so}}{3\sqrt{3}}\sum_{\langle\langle i,j\rangle\rangle{\alp}\beta}c_{i,{\alp}}^{+}v_{ij}s^{z}_{{\alp}{\beta}}c_{j,{\beta}}\nn\\
+i\frac{2{t}_{R}}{3}\sum_{\langle i,j \rangle{\alp}{\beta}}c_{i,{\alp}}^{+}\hat{z}\cdot (s_{{\alp}{\beta}}\times d_{ij})c_{j,{\beta}}+J_{\rm ex}\sum_{i,{\alp}}c_{i,{\alp}}^{+}{\bf s}\cdot{\bf M}c_{i,{\alp}}.
\label{eq:total-Hamiltonian}
\end{align}   
 where $\hat{c}_{i,{\alp}}^{+}$ ($\hat{c}_{i,{\alp}}$) creates (annihilates) an electron with spin $\alp$ on site $i$. $\langle i,j \rangle$ ($\langle\langle i,j\rangle\rangle$) runs over all the possible nearest-neighbor (next-nearest-neighbor) hopping sites. $t_{R}$ ($t_{so}$) is the Rashba (intrinsic) spin-orbit coupling constant. $v_{ij}=\pm 1$ when the trajectory of electron hopping from the site $j$ to the site $i$ is anti-clockwise (clockwise). $J_{\rm ex}$ is the ferromagnetic coupling constant. The first term denotes the nearest-neighbor hopping, the second term denotes the intrinsic spin-orbit coupling and the third one represents the extrinsic Rashba spin-orbit coupling. The fourth term is the exchange interaction between the spin of the carrier and the local moment of the ferromagnet. \par
 
We assume that the nanoribbon is uniform and periodic along the transport direction. A super unit cell is chosen as shown in the red rectangle in Fig. \ref{fig1}(a). To compute the spin torques and charge pumping, we first evaluate the nonequilibrium spin density $\delta {\bf S}$ using Kubo formula,\cite{Kurebayashi}
\begin{align}
\delta{\bf S}=&\frac{e\hbar}{2\pi A}{\rm Re}\sum_{{\bf k},a,b} \la\psi_{{\bf k}b}|\hat{\bf{s}}|\psi_{{\bf k}a}\ra\la\psi_{{\bf k}a}|  {\bf E}\cdot \hat{\bf{v}}|~\psi_{{\bf k}b}\ra\nn\\
&\times [G^{R}_{{\bf k}b}G^{A}_{{\bf k}a}-G^{R}_{{\bf k}b}G^{R}_{{\bf k}a}],
\label{eq:TSP}
\end{align}
where {\bf E} is the electric field, $\hat{\bf{v}}=\frac{1}{\hbar}\frac{\partial H}{\partial {\bf{k}}}$ is the velocity operator, $G^{R}_{{\bf k}a}=(G^{A}_{{\bf k}a})^{*}=1/(E_{F}-E_{{\bf k}a}+i\Gamma)$. $\Gamma$ is the energy spectral broadening, and $A$ is the unit cell area. $E_{F}$ is the Fermi energy, $E_{{\bf k}a}$ is the energy of electrons in band $a$. The eigenvector $|\psi_{{\bf k},a}\ra$ in band $a$ can be found by diagonalizing Eq. (\ref{eq:total-Hamiltonian}). Equation (\ref{eq:TSP}) contains both intraband ($a=b$) and interband ($a\neq b$) contributions to the nonequilibrium spin density (see the discussion in Ref. \onlinecite{Hang-2015}). The former is related to impurity scattering and the latter only includes intrinsic contributions related to Berry curvature at $\Gamma=0$. We ignore the vertex corrections as they only result in a renormalization factor of the order of unity in two dimensional hexagonal lattices\cite{Dyrdal}.

For a nanoribbon in the absence of spin-orbit couping, the eigenvalues and eigenvectors around the Dirac point are independent on the magnetization direction. However, when intrinsic spin-orbit couping is present, it acts as a valley-dependent antiferromagnetic effective field along the z direction. In the low energy limit, it reads $\sim \tau \lam_{so}\hat\sg_{z}\otimes\hat  s_{z}$. When the magnetization is directed along the x axis, the cooperation of magnetic exchange and Rashba spin-orbit coupling can open up a band gap turning the system into a (trivial) band insulator, as shown in the left panel of Fig. \ref{fig1}(b) (see also Ref. \onlinecite{Qiao3}). The corresponding density of states in the left panel of Fig. \ref{fig1}(d) displays an evident gap. In contrast, when the magnetization is directed along the z axis, the system evolves towards the quantum spin Hall regime (insulating bulk and conducting spin-polarized edges) as shown in the right panel of Fig. \ref{fig1}(b). It is related to the fact that the magnetic field couples with the intrinsic spin-orbit coupling and leads to the redistribution of ground states\cite{Kane}. Unlike the band insulator, the corresponding density of states show a parabolic dependence on energy as shown in the right panel of Fig. \ref{fig1}(d). For silicene-like materials, the exchange coupling is about 30 meV\cite{Yang1}. In this parametric range, there are only two different topological phases: trivial band insulator and quantum spin Hall insulator as shown in Fig. \ref{fig1}(c). The others topological phases such as quantum anomalous Hall insulator stand beyond this parametric range.

\begin{figure}[tbh]
\begin{center}
\includegraphics[trim=0mm 0mm 0mm 0mm,clip,scale=0.85]{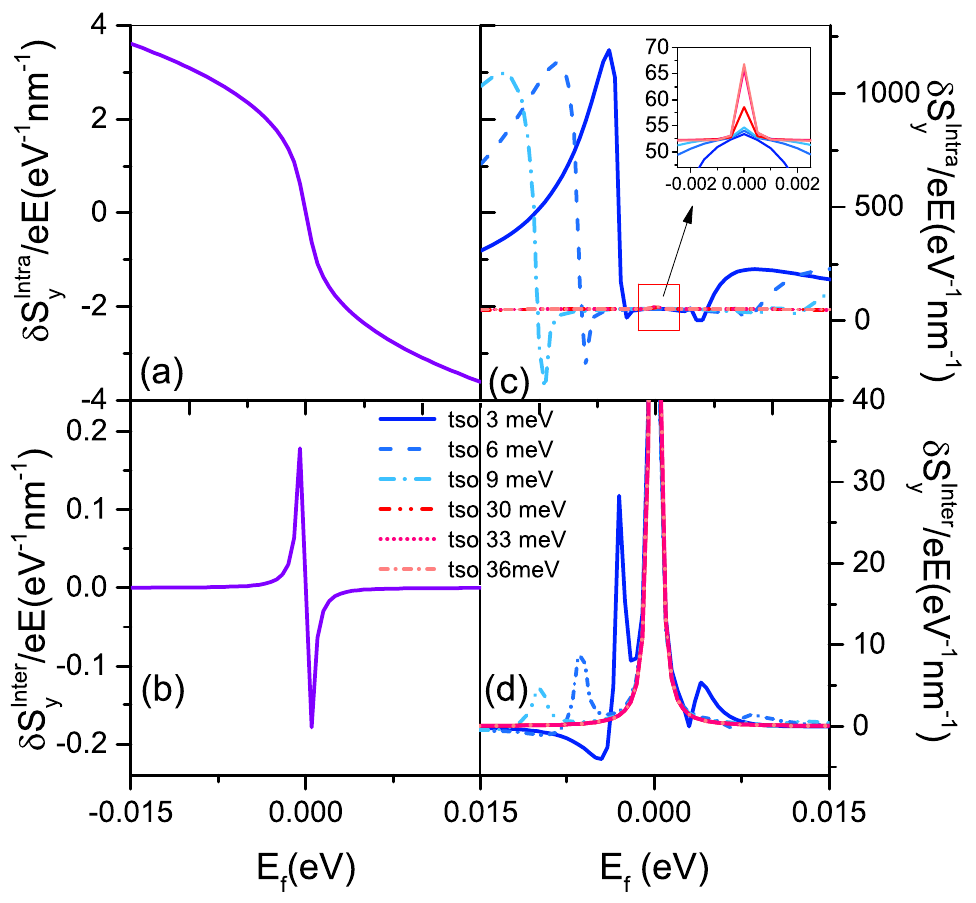}
\end{center}
\caption{(Color online) Intraband and interband components of spin density as a function of Fermi energy in a non-magnetic nanoribbon without (a)-(b) and with intrinsic spin-orbit coupling (c)-(d). The electric field is directed along the x axis. The parameters are $t_{\rm so} = 0$, $J_{\rm ex} = 0$, $t_{\rm R} = 20~meV$ and $\Gamma =0.3 ~meV$. }
\label{fig2}
\end{figure}

{\em Non-equilibrium spin density and torques} - In order to understand the influence of topological phase transition on spin-orbit torque, we first investigate the influence of intrinsic spin-orbit coupling on the nonequilibrium spin density in a non magnetic nanoribbon. In this system, Rashba spin-orbit coupling enables the electrical generation of a non-equilibrium spin density, $\delta S_y$, an effect known as the inverse spin galvanic effect and studied in details in bulk two dimensional hexagonal crystals \cite{Dyrdal,Hang-2016}. In Fig. \ref{fig2} we present the intraband (a,c) and interband contributions (b,d) to the non-equilibrium spin density in a nanoribbon as a function of Fermi energy without (a,b) and with (c,d) intrinsic spin-orbit coupling. When the intrinsic spin-orbit coupling is absent [Fig. \ref{fig2}(a,b)], the system is metallic and the intraband component dominates the spin density, indicating that carriers at the Fermi surface dominate the transport. The intraband component [Fig. \ref{fig2}(a)] is one order of magnitude larger than the interband component [Fig. \ref{fig2}(b)], in agreement with the results obtained for two-dimensional graphene-like materials, or two-dimensional electron gases\cite{Hang-apl-2013,Hang-2015,Hang-2016}. When the intrinsic spin-orbit coupling is turned on [Fig. \ref{fig2}(c,d)], it opens up a bulk band gap and induces spin-polarized edge states. In the quantum spin Hall regime (small Fermi energy, no bulk transport), the intraband and interband contributions are of the same order of magnitude, while beyond the quantum spin Hall regime (large Fermi energy, both edge and bulk transport coexist), the intraband contribution dominates the spin density.

\begin{figure}[tbh]
\begin{center}
\includegraphics[trim=0mm 0mm 0mm 0mm,clip,scale=0.85]{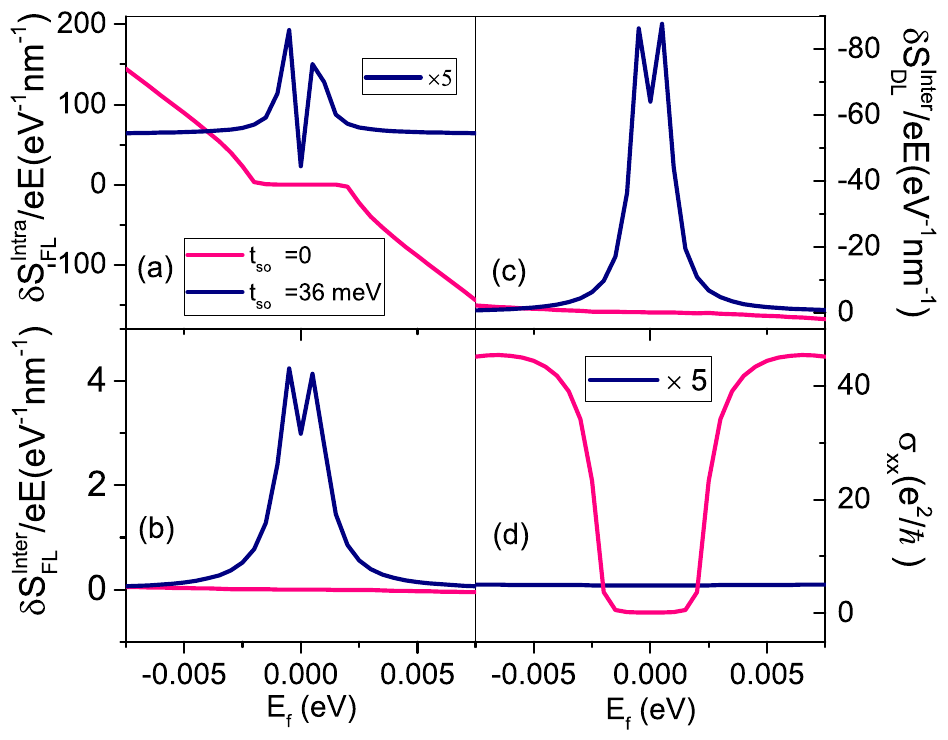}
\end{center}
\caption{(Color online)(a) Intraband, (b)-(c) interband spin density and (d) conductance as a function of Fermi energy without and with intrinsic spin-orbit coupling. The magnetization is directed along the z axis.The exchange coupling is fixed to $J_{\rm ex} = 10~meV$ and other parameters are the
same as in Fig. \ref{fig2}. }
\label{fig3}
\end{figure}

Let us now turn our attention towards the case of a magnetic nanoribbon. In our configuration, ${\bf E}=E{\bf x}$, and the non-equilibrium spin density can be parsed into two components,
\begin{align}\label{eq:spindensity}
\delta{\bf S}=\delta S_{\rm DL}{\bf y}\times{\rm\bf m}+\delta S_{\rm FL}{\bf y},
\end{align}
referred to as damping-like ($\delta S_{\rm DL}$) and field-like ($\delta S_{\rm FL}$). We plot the field-like and the damping-like spin densities with and without intrinsic spin-orbit coupling in Fig. \ref{fig3}. When the intrinsic spin-orbit coupling is absent and the exchange interaction is present, the intraband component dominates the field-like spin density in Fig. \ref{fig3}(a) and (b) similar to the case without exchange interaction displayed in Fig. \ref{fig2}(a,b). Moreover, the damping-like spin density [Fig. \ref{fig3}(b)] is smaller than the field-like spin density [Fig. \ref{fig3}(a)] because the former is a correction arising from the precession of non-equilibrium spin density around the magnetization caused by the acceleration of carriers in the electric field\cite{Sinova,Kurebayashi,Hang-2015}. When the intrinsic spin-orbit coupling is turned on, the nanoribbon enters the quantum spin Hall regime: transport only occurs through spin-polarized edge states, resulting in quantized conductance [Fig. \ref{fig3}(d)]. The interband and interband field-like spin densities [Fig. \ref{fig3}(a,c)] becomes of comparable magnitude but with opposite sign, while the damping-like spin density is significantly enhanced [Fig. \ref{fig3}(b)]. {\em As a result, the damping-like spin density dominates over the field-like spin density}. Furthermore, since the conductance is only due to edge states, the overall electrical efficiency of the torque (= torque magnitude / conductance) is dramatically enhanced in the quantum spin Hall regime.


\begin{figure}[tbh]
\centering
\includegraphics[trim = 0mm 0mm 0mm 0mm, clip, scale=0.9]{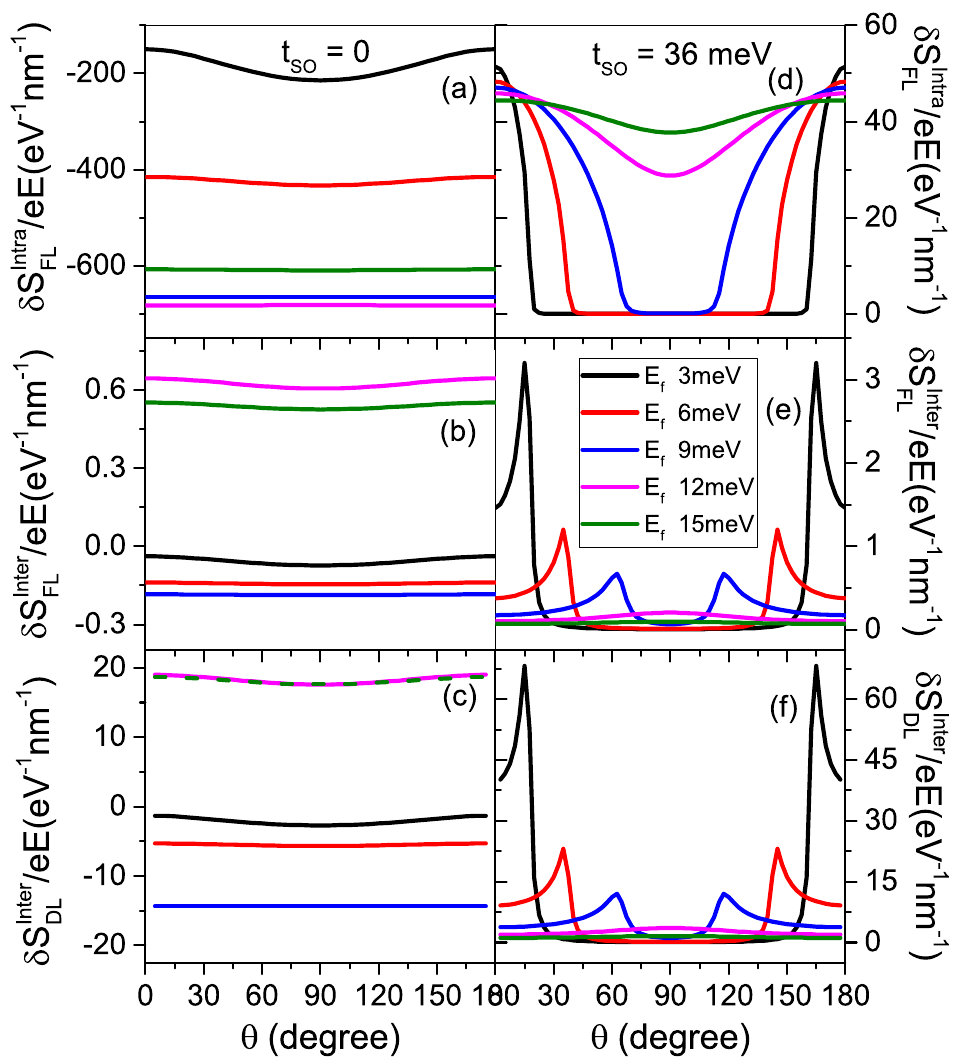}
\caption{(Color online) Intraband and interband spin density as a function of magnetization angle for different Fermi energy. (a)-(c) without intrinsic spin-orbit coupling and (d)-(f) with intrinsic spin-orbit coupling.
}
\label{fig4}
\end{figure}

The topological phase transition can be induced not only by tuning the intrinsic spin-orbit coupling but also by rotating the magnetization as shown in Fig. \ref{fig1}(b,c). In Fig. \ref{fig4}, we plot the intraband and interband contributions to spin density as a function of the magnetization angle for different Fermi energies in the absence (a,b,c) or presence (d,e,f) of intrinsic spin-orbit coupling. Dramatic features can be observed depending on whether the nanoribbon experiences a phase transition or not.\par

When intrinsic spin-orbit coupling is absent [Fig. \ref{fig4}(a,b,c)], or when intrinsic spin-orbit coupling is present and the Fermi energy large enough [$>10$  meV in Fig. \ref{fig4}(d,e,f)], the nanoribbon remains metallic independently on the magnetization direction. The spin density adopts the form given in Eq. (\ref{eq:spindensity}) and commonly observed in  two dimensional Rashba gases\cite{manchon-prb}. Minor angular dependence is observable due to the small distortion of the Fermi surface (see also Ref. \onlinecite{Lee2015}). In contrast, when intrinsic spin-orbit coupling is turned on and the Fermi energy is small enough [$<10$  meV in Fig. \ref{fig4}(d,e,f)], the nanoribbon experiences a topological phase transition from the metallic ($\theta\approx 0,\pi$) to the quantum spin Hall regime ($\theta\approx \pi/2$). This transition is clearly seen in Fig. \ref{fig4}(d), where the intraband field-like spin density decreases dramatically (but does not vanish) upon setting the magnetization away from $\theta\approx 0,\pi$. Correspondingly, the interband damping-like and field-like contributions display an abrupt and dramatic enhancement when the magnetization angle is varied through the topological phase transition. 

{\em Charge pumping} - By the virtue of Onsager reciprocity, the results obtained above for the current-driven spin densities apply straightforwardly to the charge pumping through Eq. (\ref{eq:Jcp}). From the definition of the torque, $\tau_{\rm DL}=2J_{\rm ex}\delta S_{\rm DL}/E$ and $\tau_{\rm FL}=2J_{\rm ex}\delta S_{\rm FL}/E$, and henceforth the charge current pumped by a precessing magnetization reads
\begin{align}
{\bf J}_{c}=&-\frac{2J_{\rm ex}\mu_{B}}{d\gamma}\frac{\delta S_{\rm DL}}{E}{\bf z}\times({\bf m}\times{\partial_{t}\bf m})\nn\\&+\frac{2J_{\rm ex}\mu_{B}}{d\gamma}\frac{\delta S_{\rm FL}}{E}{\bf z}\times\partial_{t}\bf m.
\end{align}
The first component gives both AC and DC signals \cite{acpumping}, while the second term is purely AC. The study of non-equilibrium spin density reported above indicates that the second component $\sim {\bf z}\times\partial_t{\bf m}$ dominates in the metallic regime (since $\delta S_{\rm FL}>\delta S_{\rm DL}$), while the first component $\sim {\bf z}\times({\bf m}\times{\partial_{t}\bf m})$ can be dramatically enhanced in the quantum spin Hall regime ($\delta S_{\rm DL}>\delta S_{\rm FL}$). Furthermore, because changing the magnetization direction can induce topological phase transitions, one expects that charge pumping with the magnetization lying out of the plane of the two dimensional nanoribbon is much more efficient than when the magnetization lies in the plane. A large charge pumping efficiency is expected at the topological phase transition. Notice though that the DC charge pumping vanishes when the magnetization precesses around the normal to the plane as $\langle{\bf m}\times{\partial_{t}\bf m}\rangle\equiv{\bf z}$.
 
{\em Discussion and conclusion} - In summary, we have investigated the impact of topological phase transition on the nature of spin-orbit torque and charge pumping in quasi-one dimensional hexagonal nanoribbons. By tuning the magnetization angle or the intrinsic spin-orbit coupling, the system can change from a band insulator to a quantum spin Hall insulator. We find that spin-charge conversion efficiencies (i.e. damping torque and charge pumping) are significantly enhanced in the quantum spin Hall regime. \par

Recently, a gigantic damping torque has been reported at the surface of topological insulators, with electrical efficiencies about two orders of magnitude larger than in transition metal bilayers \cite{Fan2014}. To the best of our knowledge, no theory is currently able to explain this observation (see discussion in Ref. \onlinecite{Theory3}). Although the present model does not precisely apply to the experimental case, it emphasizes that close or in the quantum spin Hall regime, (i) the electrical efficiency of the spin-orbit torque is dramatically enhanced due to the reduction of the conductance and, most remarkably, (ii) the competition between interband and intraband contributions reduce the field-like torque, resulting in a dominating damping-like torque. Such an effect, properly adapted to the case of topological insulators, could open interesting perspectives for the smart design of efficient spin-orbit interfaces through the manipulation of topological phase transition.\par

\begin{acknowledgments}
The research reported in this publication was supported by King Abdullah University of Science and Technology (KAUST).
\end{acknowledgments}

\end{document}